\documentclass[sigconf]{acmart}

\AtBeginDocument{%
  }

\setcopyright{acmcopyright}
\copyrightyear{2018}
\acmYear{2018}
\acmDOI{XXXXXXX.XXXXXXX}

\acmConference[Conference acronym 'XX]{Make sure to enter the correct
  conference title from your rights confirmation emai}{June 03--05,
  2018}{Woodstock, NY}
\acmPrice{15.00}
\acmISBN{978-1-4503-XXXX-X/18/06}

\usepackage{multirow} 

\usepackage{subcaption}

\begin{document}


\title{Towards Better Query Classification with Distribution-Diverse Multi-Expert Knowledge Distillation in JD Ads Search}



\author{Kun-Peng Ning}
\authornote{This work is done when Kun-Peng Ning worked at Business Growth BU, JD.COM.}
\affiliation{%
  \institution{Peking University}
  \city{Shenzhen}
  \country{China}
}
\email{ningkp@stu.pku.edu.cn}

\author{Ming Pang}
\authornote{Corresponding author.}
\affiliation{%
  \institution{Business Growth BU, JD.COM}
  \city{Beijing}
  \country{China}
}
\email{pangming8@jd.com}

\author{Zheng Fang}
\affiliation{%
  \institution{Business Growth BU, JD.COM}
  \city{Beijing}
  \country{China}
}
\email{fangzheng21@jd.com}

\author{Xue Jiang}
\affiliation{%
  \institution{Business Growth BU, JD.COM}
  \city{Beijing}
  \country{China}
}
\email{jiangxue@jd.com}

\author{Xi-Wei Zhao}
\affiliation{%
  \institution{Business Growth BU, JD.COM}
  \city{Beijing}
  \country{China}
}
\email{zhaoxiwei@jd.com}

\author{Chang-Ping Peng}
\affiliation{%
  \institution{Business Growth BU, JD.COM}
  \city{Beijing}
  \country{China}
}
\email{pengchangping@jd.com}

\author{Zhan-Gang Lin}
\affiliation{%
  \institution{Business Growth BU, JD.COM}
  \city{Beijing}
  \country{China}
}
\email{linzhangang@jd.com}

\author{Jing-He Hu}
\affiliation{%
  \institution{Business Growth BU, JD.COM}
  \city{Beijing}
  \country{China}
}
\email{hujinghe@jd.com}

\author{Jing-Ping Shao}
\affiliation{%
  \institution{Business Growth BU, JD.COM}
  \city{Beijing}
  \country{China}
}
\email{shaojingping@jd.com}

\author{Li Yuan\footnotemark[2]}
\affiliation{%
  \institution{Peking University}
  \city{Shenzhen}
  \country{China}
}
\email{yuanli-ece@pku.edu.cn}



\renewcommand{\shortauthors}{Trovato et al.}

\begin{abstract}
In the dynamic landscape of online advertising, decoding user intent remains a pivotal challenge, particularly in the context of query classification. Swift classification models, exemplified by FastText, cater to the demand for real-time responses but encounter limitations in handling intricate queries. Conversely, accuracy-centric models like BERT introduce challenges associated with increased latency.
This paper undertakes a nuanced exploration, navigating the delicate balance between efficiency and accuracy. It unveils FastText's latent potential as an 'online dictionary' for historical queries while harnessing the semantic robustness of BERT for novel and complex scenarios. The proposed Distribution-Diverse Multi-Expert (DDME) framework employs multiple teacher models trained from diverse data distributions. Through meticulous data categorization and enrichment, it elevates the classification performance across the query spectrum. Empirical results within the JD ads search system validate the superiority of our proposed approaches.

\end{abstract}



\maketitle


\section{INTRODUCTION}

Query classification plays an essential role in E-commerce search engines, which aims to understand the shopping intents of the customers from their search queries and retrieve relevant products to improve users' s satisfaction and E-commerce conversion rates \cite{boldasov2002user,chen2019fine,jiang2022short,yu2020query,liu2022pretraining}. To achieve better online generalization performance, some existing approaches \cite{nigam2019semantic,lin2020light} proposed to train a deep learning model from historical click-through data. However, as excessive model complexity, it will cause a higher online inference latency and more expensive computing costs. 


To ensure a lower online latency, some shallow models (\textit{e.g.} FastText \cite{joulin2016bag}, TextCNN \cite{chen2015convolutional}, \textit{etc.}) are widely applied in various industrial tasks because of its inference efficiency and strong memorizing ability \cite{wu2018starspace,jha2017does,tambi2020search}. Especially in E-commerce search tasks, as the massive user requests, the FastText with low online latency and stable performance become one of the most popular query classification models \cite{yu2020query,lin2018commerce,tayal2019short,park2017ebi,rygal2012properties,squizzato2015ebi,shah2022BERT}. However, the representation ability of the FastText model is insufficient, resulting the poor online generalization performance, especially on some low-frequency queries and tailed categories. Using a deeper and more complex model (\textit{e.g.} BERT \cite{devlin2018BERT}, GPT \cite{radford2018improving,radford2019language}, \textit{etc.}) is an effective solution, but it will cause a higher online inference latency and more expensive computing costs. Therefore, how to juggle both inference efficiency and online performance is obviously of great practical importance. 

\begin{figure}[t]
    \centering
    \includegraphics[width=\linewidth]{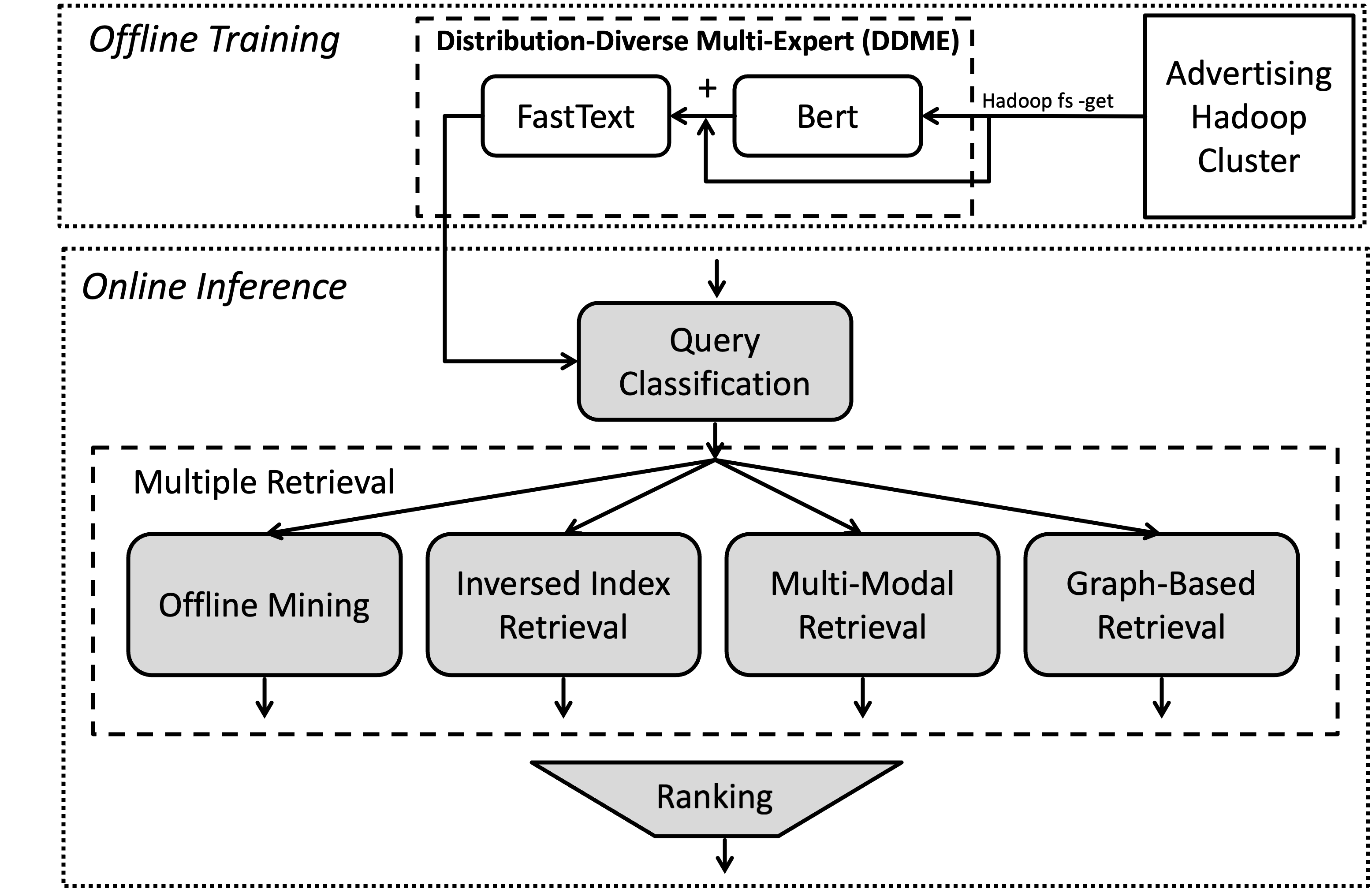}
    \caption{The overview of the search ads retrieval system based on the proposed DDME framework. In general, the whole online inference process is composed of retrieval and ranking modules. For a search query, the classifier first predicts all relevant categories for category filtering of SKU. Then, after passing the multiple retrieval module, all relevant Query-SKU pairs will be retrieved and then fed into the next ranking module. }
    \label{fig:overview}
\end{figure}

To achieve this, we start with a toy experiment to study the properties of the FastText and the BERT models in real-world JD search ads historical data. We train both models on the training set, which contains nearly 160 million user queries from historical click-through data. And we test both models on the two testing sets from T+1 and T+30 days respectively, each dataset contains nearly 2 million user queries. It is worth noting that 97.89\% of the test queries for T+1 days and 57.47\% of the ones for T+30 days have appeared in the training set, respectively. In other words, the online model has already been seen for the majority of user queries when performing T+1 day reasoning, and in fact, memorizing all these historical click-through data can also achieve good performance.
As shown in Table \ref{table:fasttext_BERT}, we found that with large parameters (17G), FastText exhibits powerful memory ability without increasing inference delay ($\sim$4ms). It is just like a huge ``online dictionary'' that performs better than BERT on search queries that have been exposed in the historical data (T+1 days). However, for some unseen queries, BERT can generalize better benefits by its stronger semantic capabilities (T+30 days). Increasing the size of parameters can also improve BERT's memory ability, which however fails to meet the maximum online delay. Obviously, combining the advantages of both is a better solution, \textit{i.e.}, employing the FastText with powerful memory ability to answer high-frequency queries, while using the BERT's semantic ability to retrieve relevant labels for unseen low-frequency queries.

By exploiting this property, we propose \textit{Distribution-Diverse Multi-Expert} (DDME) knowledge distillation, an effective Teacher-Student learning framework to boost the online search query classification performance. The student model (FastText) has the characteristics of inference efficiency, powerful memory ability, and strong robustness, but poor generalization, which is employed as an online model to ensure low inference delay. The teacher model has the characteristics of stronger generalization, but low inference efficiency, thus it can be employed as an offline model to generate high-quality training data for better the online student model training. Besides, we hope that the teacher model can perform well in queries of different frequencies. Thus, we split the historical data into high, middle, and low-frequency search queries, and then we train multiple offline BERT models from various data distributions to retrieve more potentially relevant data. As a result, more and more relevant labels not exposed in the historical data will be added to the training set and remembered by the online FastText model, while the online classification performance has been significantly improved.

\begin{table*}[t]
    \caption{The performance comparison of the FastText and BERT models on various evaluation datasets. We train both models on the training set, which has nearly 160 million user queries from historical click-through data. Then we test both models on the two testing sets from T+1 and T+30 days respectively, in which each testing set has nearly 2 million user queries. Besides, in the testing set of T+1 and T+30 days, 97.89\% and 57.47\% of test queries appeared in the training set respectively.}
    \label{table:fasttext_BERT}
    \begin{tabular}{c|c|c|c|c|c}
        \toprule
        & Parameter Size & Online Inference Latency ($\leq$20ms) & Training Set & Testing Set (T+1) & Testing Set (T+30) \\
        \hline
        \multirow{2}{*}{FastText} & \multirow{2}{*}{$\sim$17G} & \multirow{2}{*}{$\sim$4ms} & P@5: 19.2\% & \textbf{P@5: 18.7\%} & P@5: 17.3\% \\
        & & & R@5: 96.0\% & \textbf{R@5: 93.7\%} & R@5: 86.7\% \\
        \hline
        \multirow{2}{*}{BERT-base} & \multirow{2}{*}{$\sim$110M} & \multirow{2}{*}{$\sim$15ms} & P@5: 18.4\% & P@5: 18.3\% & \textbf{P@5: 18.4\%} \\
        & & & R@5: 91.5\% & R@5: 91.4\% & \textbf{R@5: 91.8\%} \\
        \hline
        \multirow{2}{*}{BERT-large} & \multirow{2}{*}{$\sim$340M} & \multirow{2}{*}{$\sim$25ms} & \multirow{2}{*}{-} & \multirow{2}{*}{-} & \multirow{2}{*}{-} \\
        & & & & &  \\
        \bottomrule
    \end{tabular}
    \vspace{0.4cm}
\end{table*}

We have developed two versions of the proposed DDME framework in the query classification task of JD search ads, each version is validated on both offline experiments and online A/B testing. For offline experiments, we conduct multiple testing datasets for offline validation, which are collected by historical click-through data and manual annotation. Besides, several evaluation metrics including precision@5 (P@5), recall@5 (R@5), and accuracy are also validated. On the other hand, for online A/B testing, we will take up 5\% of the search volume to experiment and observe the three metrics of page views (PV), item clicks (CLICK), and item pays (PAY).
Both offline experiments and online A/B testing consistently validate that, the proposed approach can significantly improve the classification performance, while achieving \textbf{+1.38\%} PV gain, \textbf{+1.33\%} CLICK gain and \textbf{+1.99\%} Pay gain, without bringing extra latency. 

The main contributions of our work are summarized as follows:
\begin{itemize}
    \item We study the properties of the FastText and the BERT models in real-world JD search ads historical data. The FastText with large model parameters exhibits powerful memory ability without increasing online inference delay, while the BERT model can generalize better for unseen queries. Combining the advantages of both can further improve the online classification performance.
    \item An effective Teacher-Student learning framework called \textit{Distribution-Diverse Multi-Expert} (DDME) knowledge distillation is proposed to boost the online performance of the FastText model under low latency constraints. It trains multiple teacher models (BERTs) from various data distributions to retrieve potential relevant data for better online student model (FastText) training, which significantly improves the online classification performance.
    \item We have developed two versions of the proposed DDME framework in the query classification task of JD search ads. Both offline experiments and online A/B testing consistently validate the effectiveness of the proposed approach.
\end{itemize}

The rest of the paper is organized as follows. We introduce the background in Section 2. Section 3 and Section 4 describe our high-level system and the detailed design of the proposed DDME framework respectively. We present offline experiments and the online A/B testing in Section 5. Related work is discussed in Section 6, and Section 7 draws the conclusion.

\section{BACKGROUND}
JD is one of the most popular E-commerce apps in China, which has more than 580 million users. JD search ads is one of the important product forms. The seller can put advertisements on some key search queries to earn more user traffic. The platform aims to recommend appropriate ads to ensure a better user experience, while optimizing the advertising strategy to maximize the ads revenue.

\begin{figure*}[t]
    \centering
    \includegraphics[width=\linewidth]{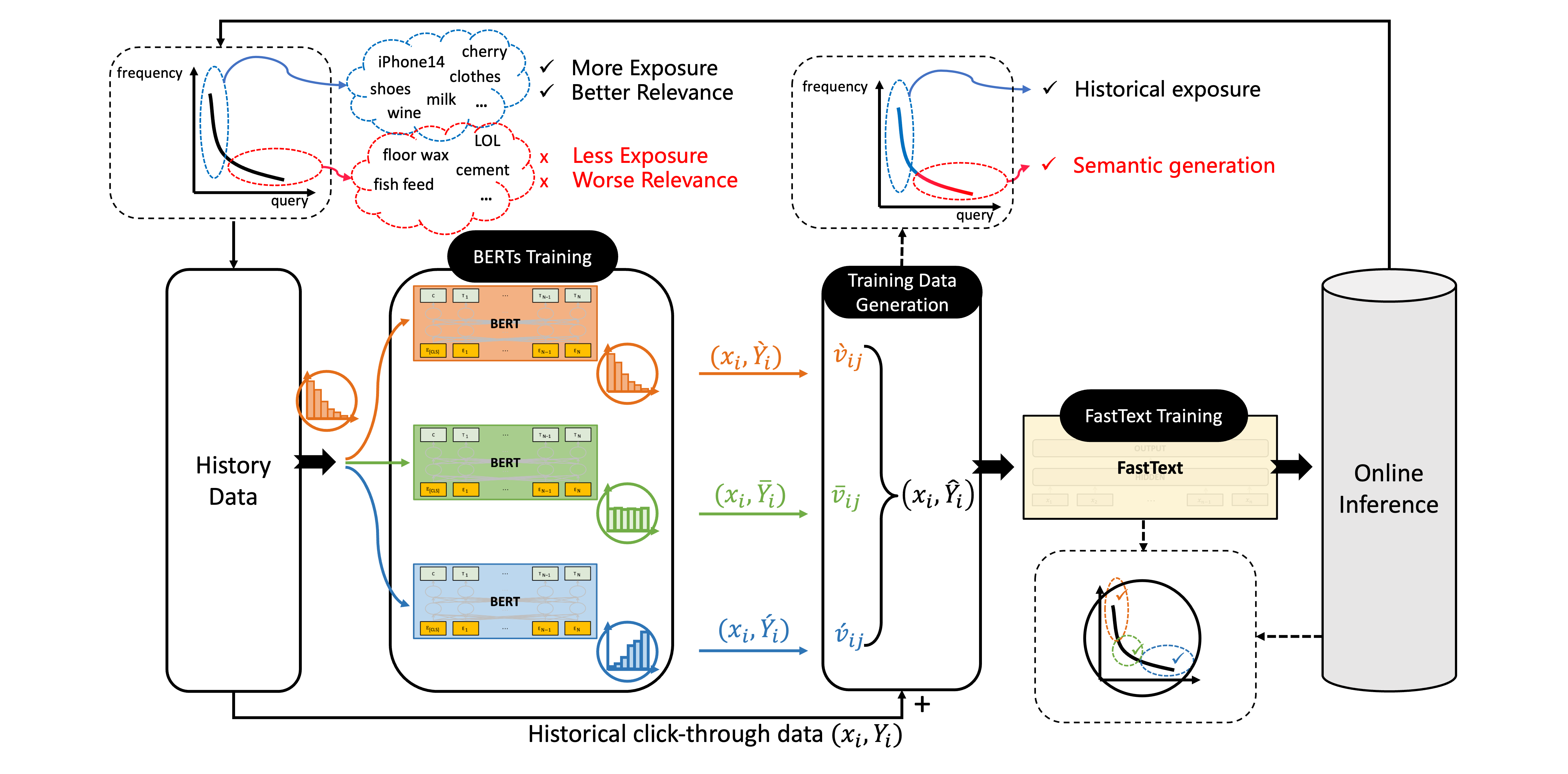}
    \caption{The proposed Distribution-Diverse Multi Expert (DDME) knowledge distillation framework in detail.}
    \vspace{0.4cm}
    \label{fig:framwork}
\end{figure*}

\section{SYSTEM OVERVIEW}
In this section, we introduce our retrieval system based on DDME from a high-level perspective and more details will be presented in the next section from a modeling perspective.

\subsection{Overview}
Figure~\ref{fig:overview} shows the overview of the search ads retrieval system based on the proposed DDME approach. In general, the whole online inference process is composed of retrieval and ranking modules. For a search query, the classifier first predicts all relevant categories for category filtering of SKU. Then, after passing the multiple retrieval module, all relevant Query-SKU pairs will be retrieved and then fed into the next ranking module. 

As the entrance of the whole search advertising system, we employ a shallow model (FastText) for efficient inference to ensure a lower online latency. However, the online performance is limited by its insufficient representation ability, especially on some low-frequency queries and tailed categories. Thus, we train multiple BERT models to retrieve more potentially relevant data for better the online FastText training. The whole process is called DDME, whose key components can be concluded into two parts: offline training, and online serving.

\subsection{Offline Training and Online Serving}
The DDME is a Teacher-Student learning framework, and the teacher (BERT) model will retrieve more relevant training data based on its strong semantic representation ability to teach the student (FastText) model to learn better. Thus, the whole process happens in the following steps.
\begin{itemize}
    \item Step 1: Multiple BERT models offline training. We train multiple BERT models using nearly 160 million historical click-through data, which updates every two days. 
    \item Step 2: Multiple BERT models offline inference. We use the learned BERT model to infer the historical data and generate the training data for the next FastText learning.
    \item Step 3: Training data generation. We merge the historical data and the generated data, and the prior distribution of the categories before and after the combination is consistent.
    \item Step 4: FastText model offline training. We train the offline FastText model using the merged data from step 3, which updates every day.
    \item Step 5: Online serving. We upload the learned FastText model for online inference.
\end{itemize}

More details can be found in Section 4.

\section{The PROPOSED APPROACH}
In this section, we first introduce some notations about the search query classification task.
Then, we investigate the properties of the FastText and the BERT models. Inspired by the observation, we finally introduce the proposed \textit{Distribution-Diverse Multi-Expert} (DDME) framework in detail.

\subsection{Notations}
The search query classification can be formulated as a multi-label classification task. Let $\mathcal{X}$ denotes the input space and $\mathcal{Y}=\{y_1,y_2,...,\\y_k\}$ be a finite set consisting of $k$ possible class labels. $\mathcal{D}=\{(x_1,Y_1),\\(x_2,Y_2),...,(x_n,Y_n)\}$ is a training set with $n$ multiple labeled instances, where $x_i\in\mathcal{X}$ is the search query and $Y_i\subseteq \mathcal{Y}$ is the multiple labels of the $i$-th example. Moreover, we denote $\mathcal{V}\in R^n\times R^k$ as the page view (PV) matrix, where $v_{ij}\in \mathcal{V}$ represents the PV of query $x_i$ and category $y_j$ pair. The goal of the query classification is to train a classifier $f:\mathcal{X}\rightarrow 2^\mathcal{Y}$ based on the multi-label training set $\mathcal{D}$.

\subsection{A Closer Look on FastText and BERT Models}
FastText has been widely used in various text classification tasks. Especially in E-commerce search tasks, benefited from its high inference efficiency, the FastText with low online latency and stable performance become one of the most popular classification models in the industry. On the other hand, the BERT model has also achieved great success in various natural language processing (NLP) tasks based on its powerful semantic representation. However, there are limited works to investigate the property of both models.

Thus we conduct toy experiments on the JD search ads dataset to explore the property of the FastText and the BERT models. Firstly, we collected the training set with nearly 160 million user queries from historical click-through data. Then, we respectively train the FastText and BERT models on this training data. Next, we collected two testing sets from T+1 and T+30 days respectively, and each testing set has nearly 2 million user queries. Importantly, in the testing set of T+1 and T+30 days, 97.89\% and 57.47\% of test queries have appeared in the training set respectively. Finally, we respectively test both models and record their performance on the T+1 and T+30 days.

Table \ref{table:fasttext_BERT} demonstrates the performance comparison of the FastText and BERT models on various evaluation datasets. It is worth noting that with large model parameters ($\sim$17G), the FastText exhibits powerful memory ability without increasing inference delay ($\sim$4ms). It can do better on the training set and testing set (T+1 day), which means that the FastText model with huge parameters has a stronger memory ability. It is just like a massive ``online dictionary'' that performs better than BERT-base on those search queries that have been exposed in the historical data. As 97.89\% of test queries on T+1 day have appeared in the training set, memorizing all these historical click-through data can indeed achieve good online performance. On the other hand, for T+30 day, only 57.47\% of test queries on T+30 day were exposed in the training set, the FastText performed worse than the BERT-base model. Benefiting from its stronger semantic capabilities, the BERT-base model can generalize better on some unseen or low-frequency search queries and retrieve more relevant data. 
Besides, despite having significantly smaller model parameters ($\sim$110M), the BERT-base model with deeper and more complex with deeper and more complex structures requires higher inference latency ($\sim$15ms), while even the online latency of the BERT-large model ($\sim$340M) cannot meet online requirements ($\leq 20ms$).

We summarized the characteristics of the two models, FastText and BERT, in the E-commerce search query classification task as follows: \textbf{the FastText with massive parameters has more powerful memory ability and faster reasoning, while the BERT with deeper and more complex structures has stronger generalization and semantic abilities.} This empirical finding motivates us to combine the advantages of both models to further improve online performance. To achieve this goal, we introduce the proposed \textit{Distribution-Diverse Multi-Expert} (DDME) knowledge distillation, a Teacher-Student learning framework illustrated in the following subsection.


\subsection{Distribution-Diverse Multi-Expert Approach}
Inspired by the empirical findings mention above, we proposed \textit{Distribution-Diverse Multi-Expert} (DDME) knowledge distillation, an effective Teacher-Student learning framework to boost the online search query classification performance. 
The student model (FastText) with massive parameters has the characteristics of inference efficiency, powerful memory ability, and strong robustness, but poor generalization on some unseen queries, which is employed as an online model to ensure low inference delay. The teacher model has the characteristics of stronger generalization, but low inference efficiency, thus it can be employed as an offline model to generate high-quality training data for better the online student model training. Besides, we hope that the teacher model can perform well in queries of different frequencies. Thus, we split the historical data into high, middle, and low-frequency search queries, and then we train multiple offline BERT models from various data distributions to retrieve more potentially relevant data. As a result, more and more relevant labels not exposed in the historical data will be added to the training set and remembered by the online FastText model, while the online classification performance has been significantly improved. 

The whole process is summarized in the DDME framework as shown in Figure \ref{fig:framwork}, which mainly composed of three components: BERTs training, training data generation, and FastText training. Specifically, we first train multiple offline BERT models from various data distributions to infer historical data. Then, we merge the inference data by the BERTs and the historical click-through data to generate the new training data for the FastText model training. This process is just like a label augmentation for this multi-label dataset. Finally, we train the FastText model based on the generated dataset and then serve for online search query classification. In the following part of this section, we will introduce these three components in detail.

\textbf{BERTs training.} As shown in Figure \ref{fig:framwork}, the historical data usually shows a long-tail distribution, in which the head (high-frequency) queries have more exposure and better query-category relevance, while the tailed (low-frequency) ones suffer from less exposure and poor relevance. In other words, these tailed queries lack enough customer behaviors information, and hence cause difficulty in identifying related categories from historical click-through log data. As a result, the learned model from these historical data still fails to perform well online on these low-frequency queries. Thus, we aim to train multiple powerful semantic models (\textit{e.g.} BERT) to retrieve more relevant data for better online model learning. 

\textit{The forward BERT expert $E_1$ training}. For a given query $x_i$, its label $y_{ij}$ and the page view (PV) $v_{ij}$ from the historical click-through data, we train the forward BERT expert $E_1$ by minimizing the following PV-reweighted binary cross-entropy (BCE) loss,
\begin{equation}
    \label{eq:E1}
    \mathcal{L}_{E_1} = -\sum_i^n \sum_j^k w_{ij}*[y_{ij} \log(\hat{y}_{ij})+(1-y_{ij})\log(1-\hat{y}_{ij})],
\end{equation}
where $w_{ij}=\frac{v_{ij}}{\sum_l^k v_{il}}$ is the PV normalized weight, $n$ and $k$ refer to the size of the dataset and labels respectively, $y_{ij}\in\{0,1\}$ indicates the ground truth of the $i$-th query is related to the $j$-th label, and $\hat{y}_{ij}\in[0,1]$ is the predicted probability. Note that it directly simulates the original long-tailed data distribution (online distribution), referring to emphasizing some high-frequency user queries.

\textit{The uniform BERT expert $E_2$ training}. To emphasize the middle-frequency user queries, we train the uniform expert by directly minimizing the BCE loss without considering PV, 
\begin{equation}
    \label{eq:E2}
    \mathcal{L}_{E_2} = -\sum_i^n \sum_j^k y_{ij} \log(\hat{y}_{ij})+(1-y_{ij})\log(1-\hat{y}_{ij}),
\end{equation}

\textit{The backward expert $E_3$ training.} We aim to train this expert to simulate the inversely long-tailed data distribution, so as to better handle tailed queries. Thus, we employ a inverse BCE loss for learning the expert $E_3$ as follows,
\begin{equation}
    \label{eq:E3}
    \mathcal{L}_{E_3} = -\sum_i^n \sum_j^k r_{ij}*[y_{ij} \log(\hat{y}_{ij})+(1-y_{ij})\log(1-\hat{y}_{ij})],
\end{equation}
where $r_{ij}=\frac{1-w_{ij}}{k-1}$ is the reverse PV normalized weight.

\textbf{Training data generation.} After learning the tree experts, more potentially relevant data from high, middle, and low-frequency queries will be retrieved. We thus employ it to infer the training data $\mathcal{D}$ and obtain the inference dataset $\bar{\mathcal{D}}$,
\begin{equation}
    \label{eq:generation2}
    \bar{\mathcal{D}} = \{(x_i,\bar{Y}_i)|\bar{Y}_i=f_{E_1}(x_i)\cup f_{E_2}(x_i) \cup f_{E_3}(x_i), (x_i,Y_i)\in D\}.
\end{equation}
where $\bar{Y}_i$ denotes the inference labels by the multiple BERTs. Then we merge the both historical data $\mathcal{D}$ and the inference data $\bar{\mathcal{D}}$ to generate the new training data $\hat{\mathcal{D}}$ for the FastText model training,
\begin{equation}
    \label{eq:generation3}
    \hat{\mathcal{D}} = \{(x_i,\hat{Y}_i)|\hat{Y}_i=\bar{Y}_i\cup Y_i, (x_i,Y_i)\in D, (x_i,\bar{Y}_i)\in \bar{\mathcal{D}}\},
\end{equation}
where $\hat{Y}_i$ is the generated labels. Intuitively, for an infrequent query $x_j$, its related categories $Y_j$ are seriously rare even missing, resulting in poor online performance on these tailed queries. Benefiting from the powerful semantic representation of the BERT model, more potentially relevant categories $\bar{Y}_i$ will be retrieved and generated into the new training set $\hat{\mathcal{D}}$ for better online FastText learning.

\textit{PV matrix $\mathcal{V}$ updating.} Obviously, how to update the PV matrix is also radical important. Many relevant labels have been retrieved by Eq \ref{eq:generation3}, and the PV matrix should also be updated accordingly. For example, for an unexposed pair of query $x_i$ and label $y_{ij}$, its original PV $v_{ij}$ is equal to $0$. Now, the BERT model retrieves the related label $y_{ij}$ and adds it into the training set, then a new problem has arisen, \textit{i.e.}, how much PV should be allocated to it. In this paper, we allocate the PV $\bar{v}_{ij}$ as follows,
\begin{equation}
    \label{eq:pv}
    \bar{v}_{ij} = p_j*\frac{M}{N_j},
\end{equation}
where 
\begin{equation*}
    p_j=\frac{\sum_{(x_i,Y_i)\in \mathcal{D}}\mathbf{1}\{j\in Y_i\}*v_{ij}}{\sum_i^n\sum_j^k v_{ij}}, \quad N_j=\sum_{(x_i,\bar{Y}_i)\in \bar{\mathcal{D}}} \mathbf{1}\{j\in \bar{Y}_i\}*v_{ij}
\end{equation*}
where $p_j$ denotes the prior probability of category $j$, $N_j$ denotes how many queries have been predicted by the multi-expert BERT models as the category $j$, and $M$ is a hyper-parameter that denotes the size of the supplementary set. Thus, the PV matrix is updated as follows,
\begin{equation}
    \hat{\mathcal{V}}=\{\hat{v}_{ij}|\hat{v}_{ij}=\mathbf{1}\{N_j>0\}*\bar{v}_{ij}+\mathbf{1}\{N_j=0\}*v_{ij}, v_{ij}\in\mathcal{V}\}.
\end{equation}
Intuitively, for a pair of query-category $(x_i,y_{ij})$, if it has appeared in the historical data, we keep its original PV $v_{ij}$. Otherwise, we allocate PV $\bar{v}_{ij}$ to make the class prior distribution unchanged.  

\textbf{FastText training.} Based on the generated training dataset $\hat{\mathcal{D}}$ and PV matrix $\hat{\mathcal{V}}$, we train the FastText model with the official version implemented by C++ code\footnote{https://github.com/facebookresearch/fastText}, and then serve for the online inference.

It is worth noting that all of these teacher model optimizations are conducted in offline experiments. The online model is completely decoupled, and its performance-improving benefits from the quality of the offline training data generation.



\begin{table}[t]
    \caption{Summary of the training and testing datasets.}
    \label{table:dataset}
    \begin{tabular}{ccccc}
        \toprule
        Dataset & Q-C pair & Query (Q) & Category (C) \\
        \hline
        training set $\mathcal{D}$ & 471861076 & 78523732 & 6440 \\
        \hline
        generated set $\bar{\mathcal{D}}$ & 775480466 & 78523732 & 6440 \\
        \hline
        \toprule
        testing set (T+1) & 50000 & 34527 & 6440 \\
        \hline
        testing set (T+30) & 50000 & 33881 & 6440 \\
        \hline
        offline annotation set & 9680 & 4500 & 6440 \\
        \hline
        online annotation set & 3000 & 3000 & 6440 \\
        \bottomrule
    \end{tabular}
\end{table}

\section{EXPERIMENTS}
In this section, we describe our experiments in detail from offline validation and online A/B testing. We first introduce the offline training data and testing data, as well as the offline evaluation metrics and compared baselines. Then we introduce our offline evaluation results, followed by the analysis of the proposed approach. Finally, we demonstrate the online A/B testing in a real-world E-commerce search engine.
\subsection{Experimental Settings}
\textbf{Training data.} For our train data $\mathcal{D}$, we sample search queries and their corresponding historical click data from the last three months of search logs. We have the frequency of click between pair of query ($x$) and category ($y$). We take the logarithm of click frequency for each $(x,y)$, which include about $471861076$ data pairs. Then, we employ the DDME approach to generate potentially relevant data, the generated set $\hat{\mathcal{D}}$ has a size of $775480466$. Table \ref{table:dataset} shows the statistics of various training sets in detail.

\textbf{Testing data.} In our offline experiments, three testing sets have been used to evaluate. Like the collection of the training set, we collect two testing sets which are divided 50 thousand pairs from historical data, and each set is respectively from T+1 and T+30 days. Moreover, we respectively collect two manual annotation sets from offline and online. Specifically, we offline collect 4500 historical search queries from high, middle, and low-frequency respectively, and employ the human oracle to annotate the ground-truth relevance of query-category pairs. Then, we can obtain a manual offline annotation set for relevance evaluation. On the other hand, when conducting online A/B testing, in addition to the impact of the algorithm on exposure, click and pay, online relevance between query and category pair should also be evaluated. Therefore, we collected the query classification results of 1500 experimental groups and 1500 control groups in the real-world E-commerce search engine, and employ humans to annotate their relevance. 
Table \ref{table:dataset} also demonstrates the statistics of various testing sets in detail.

\textbf{Offline evaluation metrics.} We introduce two metrics called P@5 and R@5 to evaluate the model classification performance on historical click-through data. The definition is as follows, 
\begin{equation}
    P@5=\frac{1}{n} \sum_i^n \frac{|\hat{Y}_i\cap Y_i|}{min(5, |\hat{Y}_i|)},
\end{equation}
\begin{equation}
    R@5=\frac{1}{n} \sum_i^n \frac{|\hat{Y}_i\cap Y_i|}{min(5,|Y_i|)},
\end{equation}
where $\hat{Y}_i$ denotes the predicted labels, and $Y_i$ denotes the ground-truth labels. Moreover, to further evaluate the performance on manual annotation sets, we introduce the accuracy with and without PV marked as ``acc'' and ``acc w/o pv'' respectively. 

\begin{table*}[t]
    \caption{The performance comparison on offline historical click-through data.}
    \label{table:offline_history}
    \begin{tabular}{c|c|c}
        \toprule
        Offline Model & Testing Set (T+1) & Testing Set (T+30)\\
        \hline
        \multirow{2}{*}{FastText (Base model)} & P@5: 19.3\% & P@5: 18.8\% \\
        & R@5: 96.6\% & R@5: 94.1\% \\
        \hline
        \multirow{2}{*}{DDME w/o Multi-Expert (Ours)}& P@5: 19.5\% & P@5: 19.1\% \\
        & R@5: 97.6\% & R@5: 95.6\% \\
        \hline
        \multirow{2}{*}{DDME (Ours)}& \textbf{P@5: 19.6\%} & \textbf{P@5: 19.3\%} \\
        & \textbf{R@5: 98.0\%} & \textbf{R@5: 96.3\%} \\
        \bottomrule
    \end{tabular}
\end{table*}

\begin{table*}[t]
    \caption{The performance comparison on offline/online manually annotated datasets.}
    \label{table:offline_annotation}
    \begin{tabular}{c|cc|cc}
        \toprule
        \multirow{2}{*}{Model} & \multicolumn{2}{c}{Offline Annotation Set} & \multicolumn{2}{c}{Online Annotation Set} \\
        & acc & acc w/o pv & acc & acc w/o pv \\
        \hline
        FastText ($\mathcal{F}$) & 92.7\% & 80.8\% & 81.5\% & 79.9\% \\
        \hline
        \textbf{DDME ($\mathcal{K}$)} & \textbf{93.2\% (+0.5\%)} & \textbf{81.5\% (+0.7\%)} & \textbf{90.6\% (+9.1\%)} & \textbf{82.1\% (+2.2\%)} \\
        \hline
        Missing Prediction ($\mathcal{F}$ - $\mathcal{F}$ $\cap$ $\mathcal{K}$) & 65.1\% & 46.2\% & - & - \\
        \hline
        \textbf{Extra Prediction ($\mathcal{K}$ - $\mathcal{K}$ $\cap$ $\mathcal{F}$)} & \textbf{73.4\% (+8.3\%)} & \textbf{56.9\% (+10.7\%)} & - & - \\
        \bottomrule
    \end{tabular}
\end{table*}

\textbf{Baselines.} To validate the effectiveness of the proposed approach, we compared the following methods in our experiments.
\begin{itemize}
    \item \textbf{FastText:} a base model for online search query classification task. It is only trained on historical click-through data.
    \item \textbf{BERT Dict:} a naive way to use the generalization ability of BERT to improve online performance. It trains an offline BERT on historical data and then generates a BERT dictionary for auxiliary online inference.
    \item \textbf{DDME w/o Multi-Expert (ours):} the degraded version of the proposed DDME approach in this paper. It only trains a single offline BERT expert $E_2$ on historical data and then generates a new training set for better online model learning.
    \item \textbf{DDME (ours):} the full version of the proposed DDME framework in this paper. By training multi-expert models from different data distributions, it can handle different frequency queries and retrieve more potentially relevant data.
\end{itemize}
It is worth noting that both versions of the proposed method use the FastText model for online search query classification. The difference is that the training set for learning the FastText model differs.

\subsection{Offline Experiments}

\begin{table*}[t]
    \caption{The results on three online A/B testings. We deployed the BERT dict method online from 2021-12-29 to 2022-01-11, and then deployed the DDME approach online from 2022-09-01 to 2022-09-07 and deployed the distribution-diverse multi-expert method from 2022-11-30 to 2022-12-06.}
    \label{table:online_AB}
    \begin{tabular}{c|c|c|c|c}
        \toprule
        Online Model & PV Gain & CLICK Gain & Pay Gain & TP99 latency \\
        \hline
        FastText (Base model) & 0\% & 0\% & 0\% & $\sim$4ms \\
        \hline
        BERT Dict + FastText (naive approach) & +0.67\% & +0.54\% & +1.03\% & $\sim$4ms \\
        \hline
        \textbf{DDME w/o Multi-Expert (ours)} & +1.10\% & +1.02\% & +1.46\% & $\sim$4ms \\
        \hline
        \textbf{DDME (ours)} & \textbf{+1.38\%} & \textbf{+1.33\%} & \textbf{+1.99\%} & $\sim$4ms \\
        \bottomrule
    \end{tabular}
\end{table*}

\begin{table*}[t]
    \caption{The performance comparison of online valid request. The online valid request indicates the number of search queries that can retrieve advertisements.}
    \label{table:online_valid}
    \begin{tabular}{c|c|c}
        \toprule
        BERT Dict + FastText (naive approach) & \textbf{DDME w/o Multi-Expert (ours)} & \textbf{DDME (ours)}\\
        \hline
        0\% & +0.29\% & \textbf{+0.52\%} \\
        \bottomrule
    \end{tabular}
\end{table*}

Table \ref{table:offline_history} and \ref{table:offline_annotation} show the offline evaluation results. We can draw the following conclusions from the results.

\textbf{Both versions of the proposed DDME can improve the classification performance of FastText.} As shown in Table \ref{table:offline_history}, we can observe that the proposed DDME approach can significantly achieve higher P@5 and R@5 performance than FastText on both testing sets of T+1 and T+30 days. The results validate that the offline BERT models can retrieve more relevant data for better FastText training. Additionally, compared to the degraded version (DDME w/o Multi-Expert), it can further improve the classification performance of FastText. It means that learning from various data distributions can further improve the classification performance of FastText.
Optimizing the retrieval ability of BERT can indirectly improve the performance of FastText. These results also validate the effectiveness of the proposed DDME framework. 

\textbf{DDME approach can indeed retrieve more relevant data.} As shown in table \ref{table:offline_annotation}, in the offline annotation set, it can be observed that the proposed DDME (denoted as $\mathcal{K}$) can achieve higher accuracy (\textbf{+0.5\% $\sim$ +0.7\%}) with(out) PV than FastText (denoted as $\mathcal{F}$). On the other hand, we validate the performance of ``missing prediction'' and ``extra prediction'', marked as $\mathcal{F}$ - $\mathcal{F}$ $\cap$ $\mathcal{K}$ and $\mathcal{K}$ - $\mathcal{K}$ $\cap$ $\mathcal{F}$ respectively. Specifically, $\mathcal{F}$ - $\mathcal{F}$ $\cap$ $\mathcal{K}$ means that will be predicted by FastText but missed by DDME. In contrast, $\mathcal{K}$ - $\mathcal{K}$ $\cap$ $\mathcal{F}$ means that will be extra predicted by DDME but FastText not. We observe that the extra prediction by the proposed DDME approach has higher accuracy (\textbf{+8.3\% $\sim$ +10.7\%}) than the missing prediction in both with(out) PV cases. In other words, more relevant data is predicted by the proposed approach while more irrelevant data is missed. 

\textbf{The DDME can improve the online classification relevance.} As shown in the online annotation set from table \ref{table:offline_annotation}, we conduct online A/B testing in the real E-commerce search engine and collect the 3000 query classification results of FastText and DDME models. After humans annotating, we observe that the proposed approach achieves significantly better classification performance (\textbf{+2.2\% $\sim$ +9.1\%}) than the FastText model. These results again validate the effectiveness of the proposed approach from another perspective.

\subsection{Online A/B Testing}

To further validate the effectiveness of the proposed DDME approach, we conducted online search query classification A/B testing in the JD search engine. We have conducted three online A/B tests in 2022 year.
Specifically, we deployed the BERT dict method online from 2021-12-29 to 2022-01-11, and then deployed the degraded version of the proposed approach (DDME w/o Multi-Expert) online from 2022-09-01 to 2022-09-07 and deployed the full version of the DDME method from 2022-11-30 to 2022-12-06. In this online A/B testing, we will take up 5\% of the search volume to experiment and observe the following metrics.

We mainly consider the following metrics: Page Views (PV), Item Clicks (CLICK), Item Pays (PAY), and TP99 latency. Where PV and CLICK represent the page views and clicks of the recommended advertisement, respectively. The PAY denotes the advertising revenue generated by user clicks. The TP99 latency, 99-th top percentile latency, is also evaluated to ensure low online delay. 
From table \ref{table:online_AB}, we can observe that the BERT dict method is a simple yet effective method to achieve various gains. However, due to its lack of flexibility, some queries are missing from the BERT dict, cause obtaining limited gain. 
Employing both versions of the proposed approach can achieve higher gains on all metrics. Especially, compared to the previous online base model, the proposed DDME approach can achieve \textbf{+1.38\%} PV gain, \textbf{+1.33\%} CLICK gain, and \textbf{+1.99\%} Pay gain, without bringing extra latency. These online results once again validate the superiority of the proposed DDME approach.

Moreover, to further validate the retrieval ability of the proposed approach, we evaluate the performance comparison of online valid requests. The online valid request indicates the number of search queries that can retrieve advertisements. As shown in Table \ref{table:online_valid}, both versions of the proposed approach in this paper can improve the performance of online valid requests. This gain means that more user search queries (\textbf{+0.52\%}) can retrieve relevant advertisements by employing the proposed DDME approach. In other words, the retrieval ability of the proposed method has been improved.

\section{RELATED WORK}
\subsection{Search Query Classification}
Search query classification plays an important role in real E-commerce search tasks \cite{shen2009product,lin2018commerce,zhang2021modeling,zhu2022enhanced,avron2021concat,ahmadvand2020jointmap,hasson2021category,nguyen2020learning,gupta2016product}. Among them, \cite{nigam2019semantic,lin2020light,sondhi2018taxonomy} proposed to employ deep learning models to improve the generalization performance. Others \cite{yu2020query,lin2018commerce,tayal2019short,park2017ebi,rygal2012properties,squizzato2015ebi,shah2022BERT} preferred to employ the FastText model to ensure online inference efficiency. To train the classifier, two approaches of collecting training data for search query classification have been proposed in \cite{shen2009product}. One is obtaining the labels of queries from click-through log data using heuristic rules, and the other is translating labeled product titles to labeled pseudo queries. Based on this, \cite{lin2018commerce,shen2009product,zhang2021modeling,nguyen2020learning,yao2021learning} also proposed using implicit feedback from user click behavior as a signal to collect training data for query classification in E-commerce tasks. In this paper, we also use click-through log data as our main training data. Moreover, we additionally generate many potentially relevant data by the offline BERT model, and then merge them into the training set for better online FastText model training.

\subsection{Knowledge Distillation}
Since \cite{hinton2015distilling} proposed knowledge distillation (KD) based on prior work \cite{ba2014deep}, KD as an effective Teacher-Student learning framework has been widely used in various tasks \cite{liu2022rethinking,jiao2019tinyBERT,papernot2016distillation,sanh2019distilBERT,touvron2021training,wang2021knowledge,jiang2020BERT2dnn}. It aims to transfer knowledge from one neural network (teacher) to another one (student) by increasing the temperature and minimizing the KL loss. 
After that, many works attempt to study its effectiveness \cite{yuan2020revisiting,phuong2019towards,tang2020understanding,cheng2020explaining,allen2020towards,stanton2021does,zhao2022decoupled,shen2021label}. Among them, \cite{stanton2021does,shen2021label} demonstrate that maintaining similar network structure (\textit{e.g.} VGG and ResNet series) has become the key to the effectiveness of the knowledge distillation. 
The proposed approach in this paper is one of the knowledge distillation frameworks, which reduces the temperature to ``condensate'' the potentially relevant labels. It is also a Teacher-Student paradigm by skillfully utilizing the property of both memorizing and generalization. 

\subsection{Long-Tailed Recognition}
In real-world search engines, the user queries always have skewed distribution with a \textit{long tail}, \textit{i.e.}, a few user queries (\textit{a.k.a.} high-frequency user queries) occupy most of the data, while most user queries (\textit{a.k.a.} low-frequency user queries) have rarely few samples \cite{stuart2010kendall, van2017devil}. In the historical click-through data, the high-frequency queries have more exposure and better query-category relevance, while the tailed ones suffer from less exposure and poor relevance \cite{zhang2021modeling,bernstein2012direct,zhu2022enhanced}. As a result, the learned model from these historical data still fails to perform well online on these low-frequency queries.
\textit{Re-sampling} methods as one of the most important data re-balancing strategies could be divided into two types: 1) Over-sampling by simply repeating data for minority classes \cite{buda2018systematic,byrd2019effect,shen2016relay}; 2) Under-sampling by abandoning data for dominant data \cite{he2009learning,japkowicz2002class}. However, duplicated tailed samples might lead to over-fitting upon minority classes, while discarding precious data will certainly impair the generalization ability of deep networks \cite{zhou2020bbn}. \textit{Re-weighting} methods are another series of prominent data re-balancing strategies, which usually allocate large weights for tail data in loss functions \cite{huang2016learning,wang2017learning}. In the search system, page view (PV) re-weighting is also commonly used to train online classification models \cite{zhu2022enhanced,zhang2021modeling}. However, for some tailed user queries, re-sampling or re-weighting these data is often ineffective, cause it lacks ground-truth relevant categories. Thus, in this paper, we propose to train offline BERT models with powerful semantic ability to generate potentially relevant data for online classifier learning.

\section{CONCLUSIONS}
In this paper, we propose a novel Teacher-Student learning framework called \textit{Distribution-Diverse Multi-Expert} (DDME) knowledge distillation to boost the classification performance of FastText under strict low online latency constraints. We first conducted toy experiments in real-world JD search ads historical data to explore the property of the FastText and the BERT models. We show that the FastText with massive parameters has more powerful memory ability and faster reasoning, while the BERT with deeper and more complex structures has stronger generalization and semantic abilities. In other words, the FastText model can do better for some user queries that have been exposed in the historical data, while the BERT can generalize better on some unseen queries benefiting from its powerful semantic representation. By combining the advantages of both, we train multiple offline BERT models from various data distributions to retrieve more potentially relevant data for better online FastText model training. As more and more relevant data not exposed in the historical data has been retrieved and added to the training set, the classification performance of the online FastText model has been significantly improved. We conduct offline experiments from multiple perspectives, and the experimental results on various datasets and metrics validate the effectiveness of the proposed approach. Moreover, we also developed two versions of the proposed DDME approach in online A/B testing. Compared to the previous base model, the proposed approach can achieve \textbf{+1.38\%} PV gain, \textbf{+1.33\%} CLICK gain and \textbf{+1.99\%} Pay gain, without bringing extra latency.

\bibliographystyle{ACM-Reference-Format}
\bibliography{www2024}

\end{document}